\documentclass[preprint,showpacs,preprintnumbers,amsmath,amssymb,nofootinbib]{revtex4-1}

\usepackage{graphicx}
\usepackage{dcolumn}
\usepackage{amssymb}
\usepackage{mathrsfs}
\usepackage{amsmath}
\usepackage{epsfig}
\usepackage{hhline}
\usepackage[utf8]{inputenc}
\usepackage{color}
\usepackage{multirow}
\usepackage[T1]{fontenc}
\usepackage{verbatim}
\usepackage{slashed}
\usepackage{epstopdf}
\usepackage{lineno,hyperref}
\usepackage{cancel}
\usepackage{xcolor}
\usepackage{fancyhdr}

\begin{document}

\title{Mono-b events from single stop production at the HL-LHC and HE-LHC}

\author{Tian-Peng Tang}
%\email{E-mail: physicspeng@outlook.com}
\author{Ning Liu}
\email{Corresponding author: liuning@njnu.edu.cn}
\author{Hang Zhou}
\email{Corresponding author: zhouhang@njnu.edu.cn}

\affiliation{Department of Physics and Institute of Theoretical Physics, Nanjing Normal University, Nanjing, 210023, China}

\begin{abstract}
Top-squarks (stop) play an important role in SUSY naturalness. The stop pair production is considered as the most effective way to search for stop at the LHC. However, the collider signature of stop pair production is usually characterized by $t\bar{t}$ plus missing transverse energy, which is also predicted in many other non-supersymmetric models. On the other hand, the single stop production via the electroweak interaction can provide some distinctive signatures, and thus will help to confirm the existence of the stop. In this paper, we investigate the observability of the mono-$b$ events from the single stop production process $pp \to \tilde t_1 \tilde{\chi}^-_1 \to b+ \slashed E_T$ in a simplified MSSM framework where the higgsinos and stops are the only sparticles at the HL-LHC and HE-LHC. We find that the stop mass and the higgsino mass may be probed up to about 1.6 TeV and 550 GeV at $5\sigma$ level at the HE-LHC with the integrated luminosity ${\cal L} = 15~\text{ab}^{-1}$. We also present the $2\sigma$ exclusion limits of the stop mass at the HL-LHC and HE-LHC.
\end{abstract}
\maketitle

\section{Introduction}

With the Higgs boson discovered in 2012~\cite{higgs-atlas,higgs-cms}, the last piece of the puzzle of the Standard Model (SM) was found. This has made the SM a great success. However, the SM is in lack of the dark matter candidate and has the hierarchy problem. Therefore, it is widely believed that the new physics will emerge at the TeV scale and stabilize the Higgs mass without fine tuning the theory's parameters. Among these models, the low-energy supersymmetry (SUSY) is one of the most promising extensions of the SM.

In supersymmetry, the top quark partners, namely stops, play a crucial role in canceling the quadratic divergence of the top quark loop, and thus protect the Higgs mass at the weak scale. In the minimal supersymmetric standard model (MSSM), the minimization conditions of the Higgs potential imply~\cite{mz}:
\begin{eqnarray}
\frac{M^2_{Z}}{2}&=&\frac{(m^2_{H_d}+\Sigma_{d})-(m^2_{H_u}+
\Sigma_{u})\tan^{2}\beta}{\tan^{2}\beta-1}-\mu^{2} \nonumber \\
&\simeq&-\mu^{2}-(m^2_{H_u}+\Sigma_{u}),
\label{minimization}
\end{eqnarray}
where the second line of Eq.(1) is valid for large $\tan\beta$. $m^2_{H_u}$ and $m^2_{H_d}$ are the weak scale soft SUSY breaking masses of the Higgs fields and $\mu$ is the higgsino mass parameter. $\tan\beta\equiv v_u/v_d$ is the ratio of vacuum expectation values of the two Higgs doublet fields $H_u$ and $H_d$. $\Sigma_{u}$ and $\Sigma_{d}$ arise from the radiative corrections to the Higgs potential and the one-loop dominant contribution to $\Sigma_{u}$ is given by \cite{tata}
\begin{eqnarray}
\Sigma_u \sim \frac{3Y_t^2}{16\pi^2}\times m^{2}_{\tilde{t}_i}
\left( \log\frac{m^{2}_{\tilde{t}_i}}{Q^2}-1\right)\,,
\label{rad-corr}
\end{eqnarray}
in which $Y_{t}$ is the Yukawa coupling of top quark and $Q$ is the renormalization scale with $Q^{2}=m_{\tilde{t}_{1}}m_{\tilde{t}_{2}}$. This indicates that only a small portion of the supersymmetric partners is closely related to the naturalness of the Higgs potential~\cite{bg}. In order to obtain the value of $M_Z$ naturally, each term in the right of Eq.~(\ref{minimization}) should be comparable in magnitude. Thus, if we now require 10\% fine tuning, the higgsino mass $\mu$ should be around $100-200$ GeV, which may be accessible by the monojet(-like) signature at the LHC~\cite{Han:2013usa}. In addition, the requirement of $\Sigma_u \sim M_Z^2/2$ leads to an upper bound on the stop mass $m_{\tilde{t}_{1,2}}\lesssim 1.5$ TeV~\cite{Baer:2012up}. Therefore, searching for the stops is a vital task to test SUSY naturalness~\cite{nsusy-1,nsusy-2,nsusy-3,nsusy-4,nsusy-7,nsusy-9,nsusy-12,Backovic:2015rwa,dutta,nsusy-14,nsusy-15,Goncalves:2016tft,Goncalves:2016nil,Wu:2018xiz,Abdughani:2018wrw,Duan:2017zar,Han:2016xet}.

During the LHC Run-1 and Run-2, the stops pair production have been extensively searched for by the ATLAS and CMS collaborations. The most effective approach to discover the stop is through the stop pair production at hadron collider. Besides the direct production, the stop quarks can also be probed via its loop effects in association with Higgs production at the LHC, especially when stops are too heavy to be directly produced~\cite{DiazCruz:2001gf}. The recent null results of LHC searches for the production of stop pair indicate that the stop mass should be heavier than 1 TeV for light neutralino~\cite{Atlas,CMS}. However, it should be emphasized that the typical collider signature of the stop pair production, such as $t\bar{t}$ plus missing transverse energy, is present in many other non-supersymmetric models. For example, in the littlest Higgs Model with $T$-parity, the pair production of $T$-odd top partner $(T_{-})$ can also give the same signature through the process $pp \to T_{-}\bar{T}_{-} \to t\bar{t}A_{H}A_{H}$, where $A_H$ is the lightest stable $T$-odd particle~\cite{Meade:2006dw,Han:2008gy,Yang:2018oek}. Therefore, if such a signature is observed in the pair production channel in future experiments, we still need other information to identify the existence of the stop. Like the study of top quark, one possible way is to further look for the single production of stop via electroweak interaction (see Fig.~\ref{fig:feynman_diagram}), which can provide some unique signatures at the LHC~\cite{single-stop-1,single-stop-2,wu-1,wu-2}.

Besides the LHC, the High Luminosity LHC (HL-LHC) and High Energy LHC (HE-LHC) have been widely discussed. The former will run at a colliding energy $\sqrt{s}=14$ TeV with the integrated luminosity of 3 ab$^{-1}$, while the latter will be designed to operate at a center of mass energy $\sqrt{s}=27$ TeV with the integrated luminosity of 15 ab$^{-1}$ over 20 years of operation. In this work, we investigate the single stop production process $pp \to \tilde{t}_1 \tilde{\chi}^{-}_{1}$ in a simplified framework where the higgsinos and stops are the only sparticles in the MSSM at the HL-LHC and HE-LHC. Such a scenario is favored by the natural SUSY and has been widely studied in~\cite{snsusy-1,snsusy-2,snsusy-3,snsusy-4,snsusy-5}. It should be noted that the thermal relic density of the light higgsino-like neutralino dark matter is typically low because of the large annihilation rate in the early universe~\cite{Abdughani:2019wss,Abdughani:2019wai,Abdughani:2017dqs,Wu:2017kgr,Kobakhidze:2016mdx,Han:2014xoa,Cao:2013mqa}. This leads to the standard thermally produced WIMP dark matter inadequate in the natural SUSY. In order to provide the required relic density, some alternative ways have been proposed~\cite{Acharya:2010af,Moroi:1999zb,Gelmini:2006pw,Acharya:2007rc,Acharya:2008zi,Choi:2008zq}, such as the axion-higgsino admixture as the dark matter~\cite{Baer:2011hx,Baer:2011uz}. Due to the higgsinos being nearly degenerate, their decay products are very soft so that they will mimic the missing energy at the LHC. Hence the single stop production $pp \to \tilde{t}_1 \tilde{\chi}^{-}_{1}$ will give two distinctive signatures, namely, the mono-$t$ events from $\tilde{t}_1 \to t \tilde{\chi}^0_{1,2}$ and the mono-$b$ events from $\tilde{t}_{1} \to b \tilde{\chi}^{+}_{1}$. The sensitivity of the hadronic and leptonic mono-$t$ events has been studied at the HL-LHC in Ref.~\cite{wu-1}.
% Besides the HL-LHC, the HE-LHC is designed to operate at a center of mass energy $\sqrt{s}=27$ TeV with the integrated luminosity of 15 ab$^{-1}$ over 20 years of operation.
We will focus on the mono-$b$ analysis and explore its observability at the HL-LHC and HE-LHC. % The discovery reach for TeV sparticles the HE-LHC has been studied in [].

This paper is organized as follows. In Sec.~II, we calculate the cross section of the single stop electroweak production process $pp \to \tilde{t}_1 \tilde{\chi}^{-}_{1}$. Then in Sec.~III, we perform Monte Carlo study of the mono-$b$ signature from this single stop production at the HL-LHC and HE-LHC. Finally in Sec.~IV, we draw our conclusions.

\section{Calculation of single stop production}

In the MSSM, the kinetic terms of top-squark are given by
\begin{alignat}{5}
 \mathcal{L} = \sum_{\tilde{t}}\,(\partial_\mu \tilde{t}^*_L \, \partial_\mu\tilde{t}^*_R)\,\left(
\begin{array}{c}
 \partial^\mu\,\tilde{t}_L \\ \partial^\mu\,\tilde{t}_R
\end{array}
 \right) - (\tilde{t}^*_L \, \tilde{t}_R^*)\,M_{\tilde{t}}^2\,
\left(
\begin{array}{c}
\tilde{t}_L \\ \tilde{t}_R
\end{array}
 \right),
\label{eq:stop}
\end{alignat}
with the stop mass matrix
\begin{equation}
 {M}^2_{\tilde{t}} =
\left(
\begin{array}{cc}
m^{2}_{\tilde{Q}_{3L}}+m^{2}_{t}+D^{t}_L & m_{t}X^{\dag}_{t} \\
m_{t}X_{t} & m^{2}_{\tilde{U}_{3R}}+m^{2}_{t}+D^{t}_R
\end{array}
\right),
\label{eq:stopmass}
\end{equation}
where
\begin{eqnarray}
D^{t}_L=m^{2}_{Z}(\frac{1}{2}-\frac{2}{3}\sin^{2}\theta_{W})\cos2\beta,  \quad D^{t}_R=\frac{2}{3}m^{2}_{Z}\sin^{2}\theta_{W}\cos2\beta, \quad X_{t}=A_{t}-{\mu}{\cot\beta}.
\end{eqnarray}
Here $m_{\tilde{Q}_{3L}}$ and $m_{\tilde{U}_{3R}}$ are the soft SUSY-breaking mass parameters, and $A_{t}$ is the trilinear coupling. $\mu$ is the higgsino mass parameter. The mass eigenstates $\tilde t_1$ and $\tilde t_2$ can be obtained by a unitary transformation,
\begin{eqnarray}
\begin{pmatrix} \tilde t_1\\ \tilde t_2 \end{pmatrix} &=& \begin{pmatrix} \cos\theta_{\tilde{t}}&& \sin\theta_{\tilde{t}}\\ -\sin\theta_{\tilde{t}} && \cos\theta_{\tilde{t}} \end{pmatrix} \begin{pmatrix} \tilde t_L\\ \tilde t_R \end{pmatrix},
\end{eqnarray}
where the mixing angle $\theta_{\tilde t}$ is given by $\sin 2\theta_{\tilde t} = \frac{2m_t X_t}{m_{\tilde{t}_1}^2-m_{\tilde{t}_2}^2}$ and $\cos 2\theta_{\tilde t} = \frac{m_{\tilde{Q}_{3L}}^2+D_L^t-m_{\tilde{U}_3{R}}^2-D_R^t}{m_{\tilde{t}_1}^2-m_{\tilde{t}_2}^2}$.

Besides, the mass matrix of the neutralinos $\tilde{\chi}_{1,2,3,4}^0$ in gauge-eigenstate basis $(\tilde{B}, \tilde{W}, \tilde{H}_d^0, \tilde{H}_u^0)$ is given by
  \begin{eqnarray}
  \label{mass}
  M_{\chi^0}=
\left(
   \begin{array}{cccc}
     M_1 &0&-\cos\beta \sin\theta_W m_Z&\sin\beta \sin\theta_W m_Z\\
     0& M_2&\cos\beta \cos\theta_W m_Z&\sin\beta \cos\theta_W m_Z\\
     -\cos\beta \sin\theta_W m_Z&\cos\beta \cos\theta_W m_Z&0&-\mu\\
     \sin\beta \sin\theta_W m_Z&-\sin\beta \cos\theta_W m_Z&-\mu&0\\
     \end{array}
 \right)
\end{eqnarray}
where $M_{1,2}$ are soft-breaking mass parameters for bino and wino. Eq.~\ref{mass} can be diagonalized by a unitary $4\times 4$ matrix $N$~\cite{matrix}. While the mass matrix of the charginos $\tilde{\chi}_{1,2}^\pm$ in the gauge-eigenstates basis ($\tilde{W}^+, \tilde{H}_u^+$, $\tilde{W}^-, \tilde{H}_d^-$) is given by
 \begin{eqnarray}
  \mathrm{M}_{\chi^{\pm}}=
\left(
   \begin{array}{cc}
     0 &X^T\\
     X& 0\\
     \end{array}
 \right)
\end{eqnarray}
with
\begin{eqnarray}
  X=
\left(
   \begin{array}{cc}
     M_2 &\sqrt{2}s_{\beta}m_W\\
     \sqrt{2}c_{\beta}m_W& \mu\\
     \end{array}
 \right).
\end{eqnarray}
Here the mass matrix $X$ can be diagonalized by two unitary $2\times 2$ matrices $U$ and $V$~\cite{matrix}.

When $m_{U3R} \ll m_{Q3L}$ and $\mu \ll M_{1,2}$, the lighter stop $\tilde{t}_1$ is dominated by right-handed stop component and the electroweakinos $\tilde{\chi}^0_{1,2}$ and $\tilde{\chi}^\pm_1$ are higgsino-like. Then $\tilde{t}_1$ will mainly decay to $b\tilde{\chi}^+_1$ with the branching ratio $\sim 50\%$. As a proof of concept, we focus on a simplified MSSM framework where the higgsinos and right-handed stop are the only sparticles in our following study.

%%%--Feynman diagrams--%%%
\begin{figure}[htbp]
\small
\centering
\includegraphics[width=10cm,height=3cm]{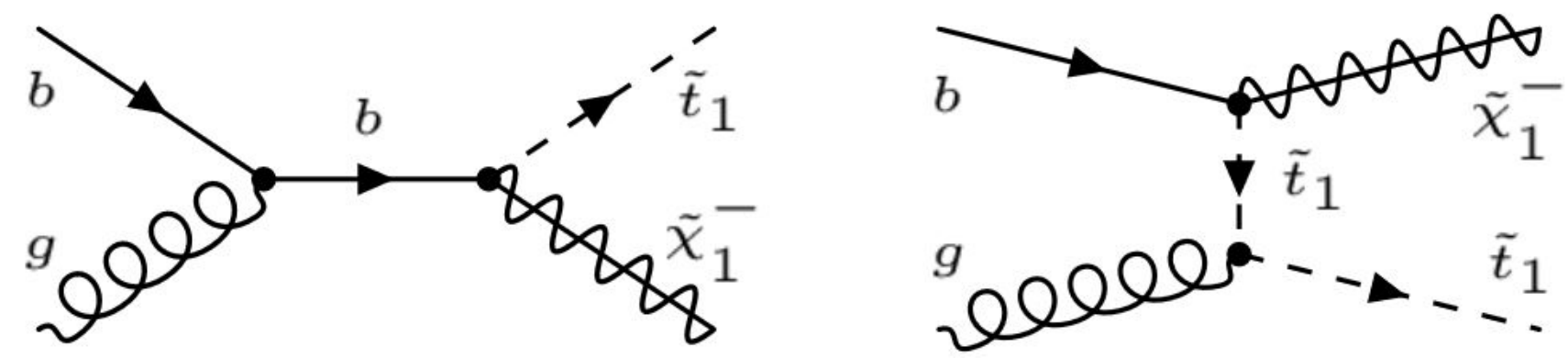}
\caption{Feynman diagrams of the single stop production process $pp \to \tilde{t}_1 \tilde{\chi}^{-}_{1}$ at the partonic level.}
\label{fig:feynman_diagram}
\end{figure}
%%%--end--%%%
%In addition, the stop electroweak interaction to LO is\cite {electroweak}:
%
%\begin{eqnarray}
%g_{\tilde t_a b \tilde{\chi}^{-}_i} &=& \tilde{t}_a\,\bar{b}\,\left(g^{(1)}_{\tilde{t} \tilde{\chi}^{-}_i b, R}P_R
%+ g^{(2)}_{\tilde{t} \tilde{\chi}^{-}_i b, L}P_L\right)\,\tilde{\chi}_i^-,\\
%g_{\tilde t_a b \tilde{\chi}^{+}_i} &=& \tilde{t^*_a}\,\tilde{\chi}_i^+\,\left(g^{(1)}_{\tilde{t^*} \tilde{\chi}^{+}_i b, R}P_R
%+ g^{(2)}_{\tilde{t^*} \tilde{\chi}^{+}_i b, L}P_L\right)b,
%\label{eq:stopcha-lo}
%\end{eqnarray}
%
%with the coupling
%
%\begin{eqnarray}
%g^{(a)}_{\tilde t_1 b \tilde{\chi}^{-}_i} &=& R^{\tilde t}_{11}g^{(a)}_{\tilde{t} \tilde{\chi}^{-}_i b, L} + R^{\tilde t}_{12}g^{(a)}_{\tilde{t} \tilde{\chi}^{-}_i b, R},\\
%g^{(a)}_{\tilde t^*_1 b \tilde{\chi}^{-}_i} &=& R^{\tilde t}_{11}g^{(a)}_{\tilde{t^*} \tilde{\chi}^{+}_i b, L} + R^{\tilde t}_{12}g^{(a)}_{\tilde{t^*} \tilde{\chi}^{+}_i b, R},
%\end{eqnarray}
%
%and the projection operators
%
%\begin{eqnarray}
%P_{L,R} &=& \frac{1}{2}\left(1\mp\gamma^{5}\right).
%\end{eqnarray}
%
%Meanwhile, the coupling of $g\tilde{t}\tilde{t}$ can be written by,
%\begin{eqnarray}
%\mathcal{L}_{g\tilde{q}\bar{\tilde{q}}} = -ig_s\left(\tilde{q}^\dag\frac{\lambda_A}{2}\partial_\mu\tilde{q} - \partial_\mu\tilde{q}^\dag\frac{\lambda_A}{2}\tilde{q}\right)G^\mu_A.
%\end{eqnarray}
In Fig.~\ref{fig:feynman_diagram}, we show the Feynman diagrams of the single stop production process $g(p_a)b(p_b) \to \tilde{t}_1(p_1) \tilde{\chi}^{-}_{1}(p_2)$, whose amplitudes are given by,
\begin{eqnarray}
i\mathcal{M}^{(s)}&=&g_s g_{eff}T^a_{\alpha\beta}\frac{\bar{u}(p_2)(\sin\theta_{\textrm{eff}}P_R+\cos\theta_{eff}P_L)\gamma_\mu u(p_b)}{(\slashed p_a+\slashed p_b)-m_b+i\epsilon}\epsilon^\mu(p_a),\\
i\mathcal{M}^{(t)}&=&g_sg_{eff}T^a_{\alpha\beta}\frac{\bar{u}(p_2)(\sin\theta_{\textrm{eff}}P_R+\cos\theta_{eff}P_L)u(p_b)}{(p_b-p_2)^2-m_{\tilde{t}_1}^2+i\epsilon}
(p_{1\mu}-p_{b\mu}+p_{2\mu})\epsilon^{\mu}(p_a),
\end{eqnarray}
with
\begin{eqnarray}
\tan\theta_{\textrm{eff}}=\frac{y_{b}U^{*}_{12}\sin\theta_{\tilde{t}}}{-g_2 V_{11}\cos\theta_{\tilde{t}}+y_{t}V_{12}\sin\theta_{\tilde{t}}}.
\end{eqnarray}
Here $y_b$ and $y_t$ are the bottom and top quark Yukawa coupling, respectively. $T_a$ are the Gellman-matrices. %Then, neglecting the bottom mass $m_b$ and keeping the bottom Yukawa coupling $y_b$, we calculate the squared matrix element of the process ,
%
%\begin{eqnarray}
%\begin{aligned}
%\overline{|\mathcal{M}|}^{2}=&{\frac{2}{3}g_{\textrm{eff}}^2g_s^2\Bigl[\frac{2E_{cm}^2(m_1^2-m_2^2-E_{cm}^2)+4m_2^2\sin{2\theta_\textrm{eff}}(2m_2^2-E_{cm}^2)+4E_{cm}^3|\vec{p}_1|\cos\theta}{E_{cm}^2(E_{cm}^2+m_1^2-m_2^2-2E_{cm}|\vec{p}_1|\cos\theta)}}\\
%&{+\frac{E_{cm}^2+m_2^2-m_1^2+2E_{cm}|\vec{p}_1|\cos\theta}{E_{cm}^2}}\Bigl],
%\end{aligned}
%\end{eqnarray}
%
Then, we can have the partonic cross section in the center-of-mass frame at the leading order,
\begin{eqnarray}
\frac{d\hat\sigma}{d\cos\theta}=\frac{|\vec{p}_1|}{16\pi \hat{s}^{3/2}}\overline{|\mathcal{M}|}^{2}
\end{eqnarray}
where
\begin{eqnarray}
|\vec{p}_1|^2=\frac{(\hat{s}-m_{\tilde{t}_1}^2-m_{\tilde{\chi}^-_1}^2)^2-4m_{\tilde{t}_1}^2m_{\tilde{\chi}^-_1}^2}{4\hat{s}}, \quad \hat{s}=(p_a+p_b)^2.
\end{eqnarray}
%and $\theta$ is the angle of $\vec p_a$ and $\vec p_1$.
%
%Thus, the integrated partonic cross section behaves as,
%\begin{eqnarray}
%\begin{aligned}
%\hat\sigma=&\frac{g_{\textrm{eff}}^2g_s^2}{12\pi}\Bigl[\frac{2E_{cm}^3|\vec p_1|+E_{cm}^2(m_1^2-m_2^2)-m_2^2\sin{2\theta_{\textrm{eff}}}(E_{cm}^2-2m_2^2)}{E_{cm}^6}\\
%&\times\log(\frac{E_{cm}^2+m_1^2-m_2^2+2E_{cm}|\vec p_1|}{E_{cm}^2+m_1^2-m_2^2-2E_{cm}|\vec p_1|})+\frac{(E_{cm}^2+m_2^2-m_1^2)|\vec p_1|}{E_{cm}^5}\Bigl].
%\end{aligned}
%\end{eqnarray}
Subsequently, the corresponding hadronic cross section can be obtained by convoluting the partonic cross section $\hat{\sigma}({gb \to \tilde{t}_1 \tilde{\chi}^{-}_{1}})$ with the parton distribution functions(PDFs), which is given by,
\begin{eqnarray}
\sigma=\sum_{b,g} \int dx_1 f_{b/p}(x_1, \mu_F^2) \int dx_2 f_{g/p}(x_2, \mu_F^2) \hat\sigma_{gb \to \tilde{t}_1 \tilde{\chi}^{-}_{1}}(x_1 x_2 s).
\end{eqnarray}
where $s$ is the squared $pp$ centre-of-mass energy. The PDF $f_{b/p}$ is the number density of bottom quark carrying a fraction $x_1$ of the momentum of the first proton, and similarly with the PDF $f_{g/p}$ for the other proton. In our calculations, we use the CTEQ6L set with the factorization scale $\mu_F$ and renormalization scale $\mu_R$ chosen to be $\mu_R=\mu_F=m_Z$. We calculate the leading order (LO) cross sections of the process $pp \to \tilde{t}_1\tilde{\chi}^+_1$ with the \textsf{MadGraph5\_aMC@NLO}~\cite{Madgraph} and include the NLO QCD corrections by applying a $K$ factor of 1.4~\cite{Jin:2003ez,Jin:2002nu,electroweak}. The NLO QCD corrected cross sections of the process $pp \to \tilde{t}_1\tilde{t}^*_1$ with the \textsf{Prospino}~\cite{Beenakker:1996ed}.

%%%--cross sections--%%%
\begin{figure}[htbp]
%\small
\centering
\includegraphics[width=14cm,height=10cm]{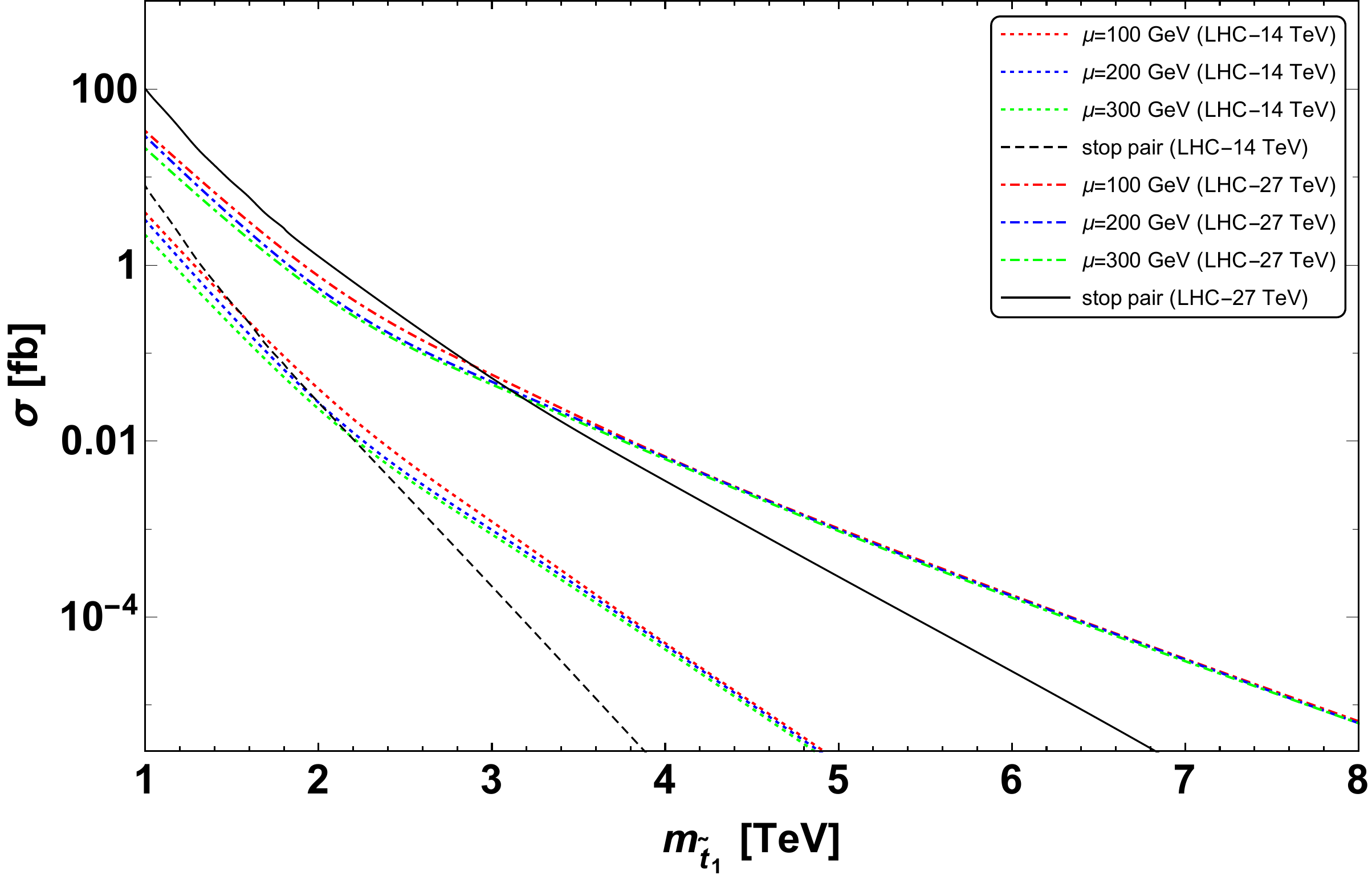}
\vspace{-0.5cm}
\caption{The hadronic
 cross sections of the stop pair production process $pp \to \tilde{t}_1\tilde{t}^*_1$ and the single stop production process $pp \to \tilde{t}_{1}\tilde{\chi}^{-}_{1}$ at the 14 TeV and 27 TeV LHC. The contribution of the charge-conjugate process of the single stop production $pp \to \tilde{t}^*_{1}\tilde{\chi}^{+}_{1}$ is included.}
\label{fig:section}
\end{figure}
%%%--end--%%%
In Fig.~\ref{fig:section}, we present the cross sections of the processes $pp \to \tilde{t}_1\tilde{t}^*_1$ and $pp \to \tilde{t}_{1}\tilde{\chi}^{-}_{1}$ at the LHC and HE-LHC. Since the electroweakinos can be still light~\cite{Athron:2018vxy}, we take the higgsino mass parameter $\mu=100,200,300$ GeV as examples. We vary the higgsino mass parameter $\mu$ and right-handed stop soft mass $m_{U3}$, and fix other soft supersymmetric masses at 1 TeV. We use the package \textsf{SUSYHIT}~\cite{SUSYHIT} to calculate masses, couplings and branching ratios of the sparticles. From Fig.~\ref{fig:section}, we can see that the cross section of the single stop production decreases slower than that of the stop pair production as stop becomes heavy. This is particular interesting for a stop in the TeV region. When the stop is heavier than about 2.2 TeV (for 14 TeV LHC) and 3.3 TeV (for 27 TeV LHC), the single stop production may have a larger production rate than the stop pair production, due to the larger phase space.

\section{Observability of mono-bottom signature at the HL/HE-LHC}
Next, we investigate the mono-$b$ signature for the single stop production, which is given by
\begin{eqnarray}
pp \to \tilde{t}_1 \tilde{\chi}^{-}_{1} \to b \tilde{\chi}^+_{1} \tilde{\chi}^-_1 \to b+ \slashed E_T.\label{monob}
\end{eqnarray}
It should be noted that the chargino $\tilde{\chi}^\pm_1$ in our scenario is treated as $\slashed E_T$ because the mass difference between it and the LSP neutralino is small so that their decay products are very soft in the detectors. We use \textsf{MadGraph5\_aMC@NLO} to generate the parton-level signal and background events. The parton shower and the detector simulations are implemented by \textsf{Pythia}~\cite{pythia} and \textsf{Delphes}~\cite{delphes} within the framework of \textsf{CheckMATE2}~\cite{checkmate}. We use the anti-$k_t$ jet clustering algorithm with a radius parameter $R=0.4$~\cite{anti-kt} and assume the $b$-jet tagging efficiency and mistagging efficiency as 80\% and 0.2\%, respectively. The largest SM background comes from the process $Z+jets$ because the light-flavor jets can be mis-tagged as $b$-jets. The subdominant backgrounds are the semi- and full-hadronic $t\bar t$ due to the mis-measurement of $\slashed E_T$ from the undetected leptons and the limited jet energy resolution.
%%%--distribution_14TeV--%%%
\begin{figure}[htbp]
%\small
\centering
\includegraphics[width=8cm,height=9cm]{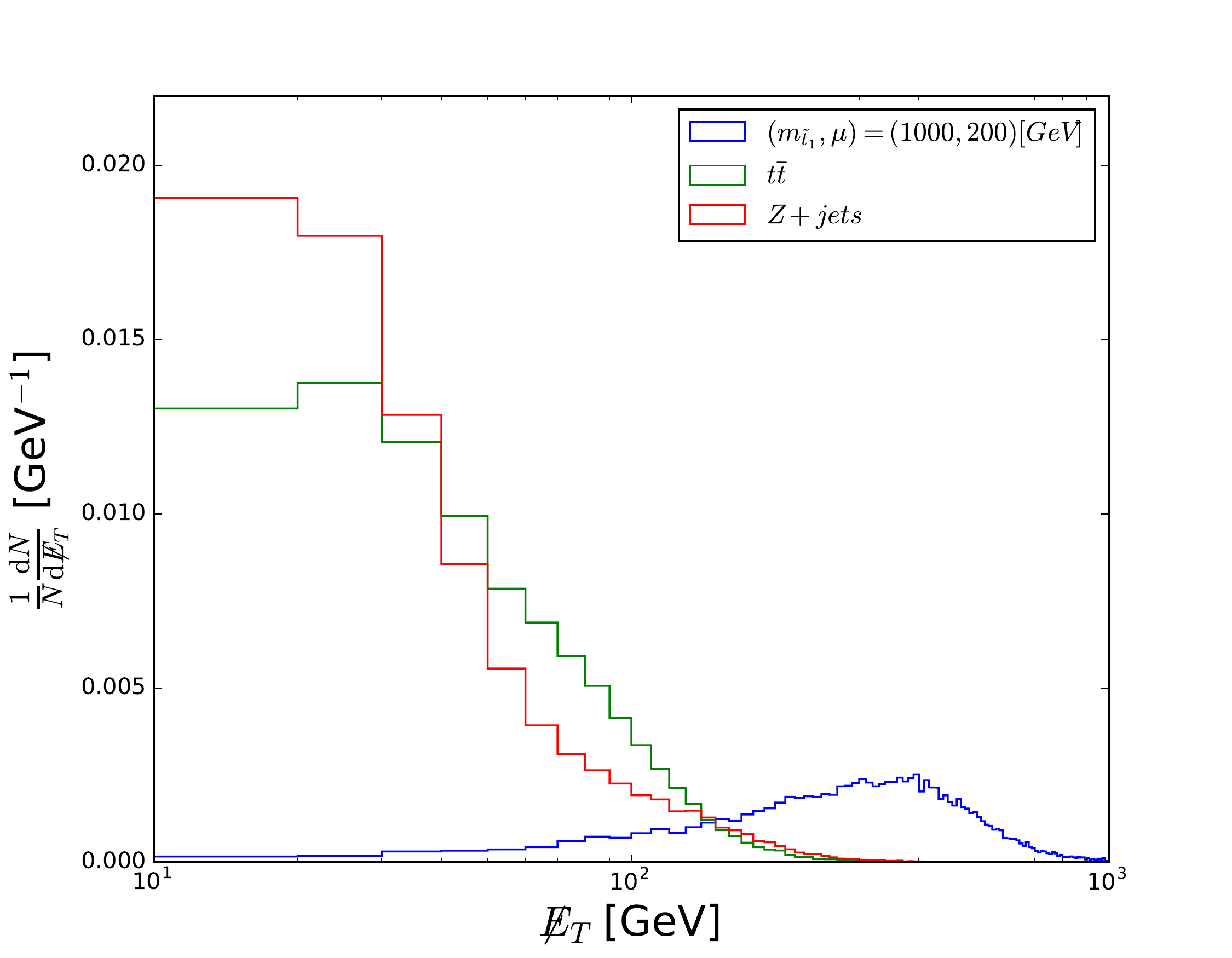}
\includegraphics[width=8cm,height=9cm]{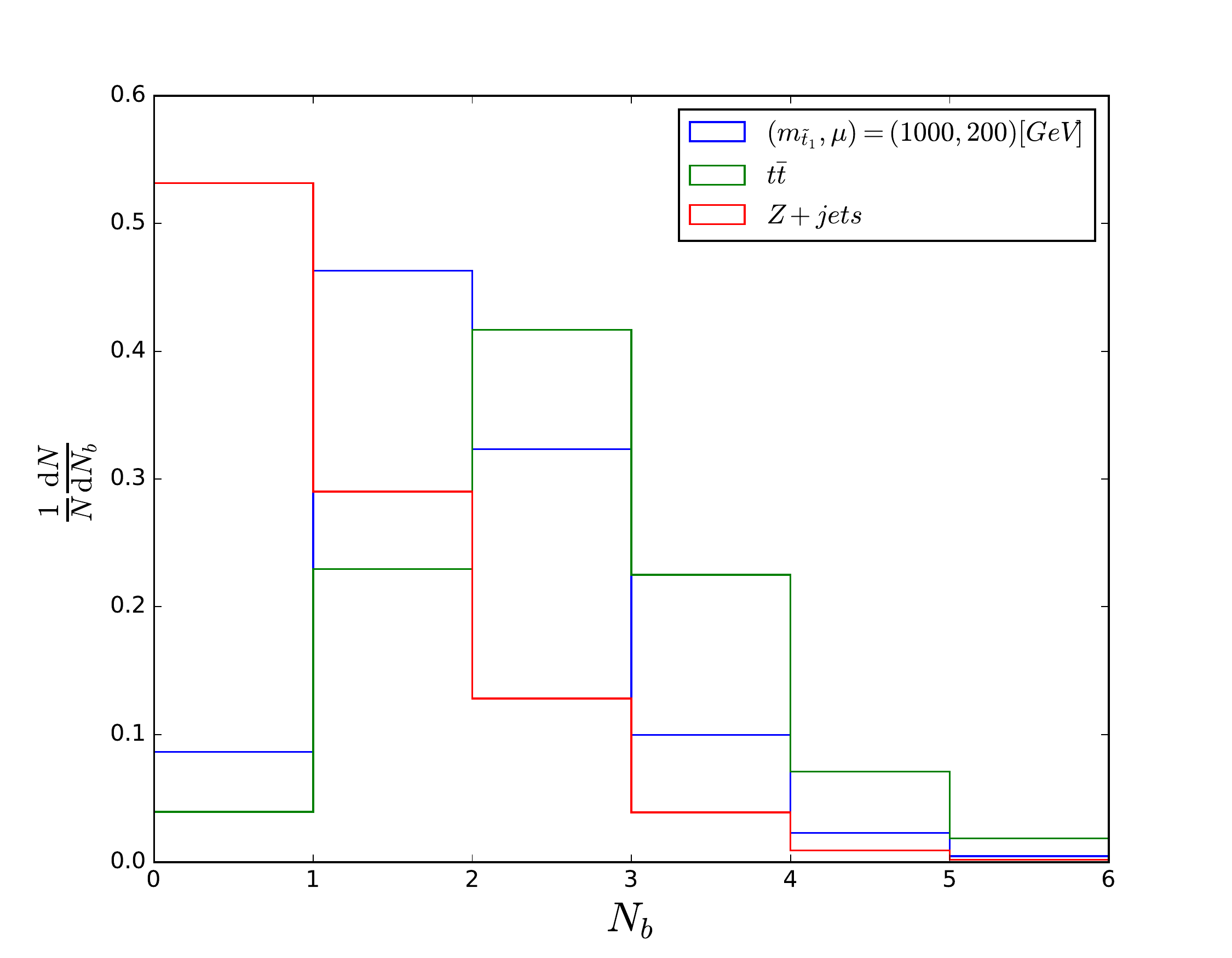}
\includegraphics[width=8cm,height=9cm]{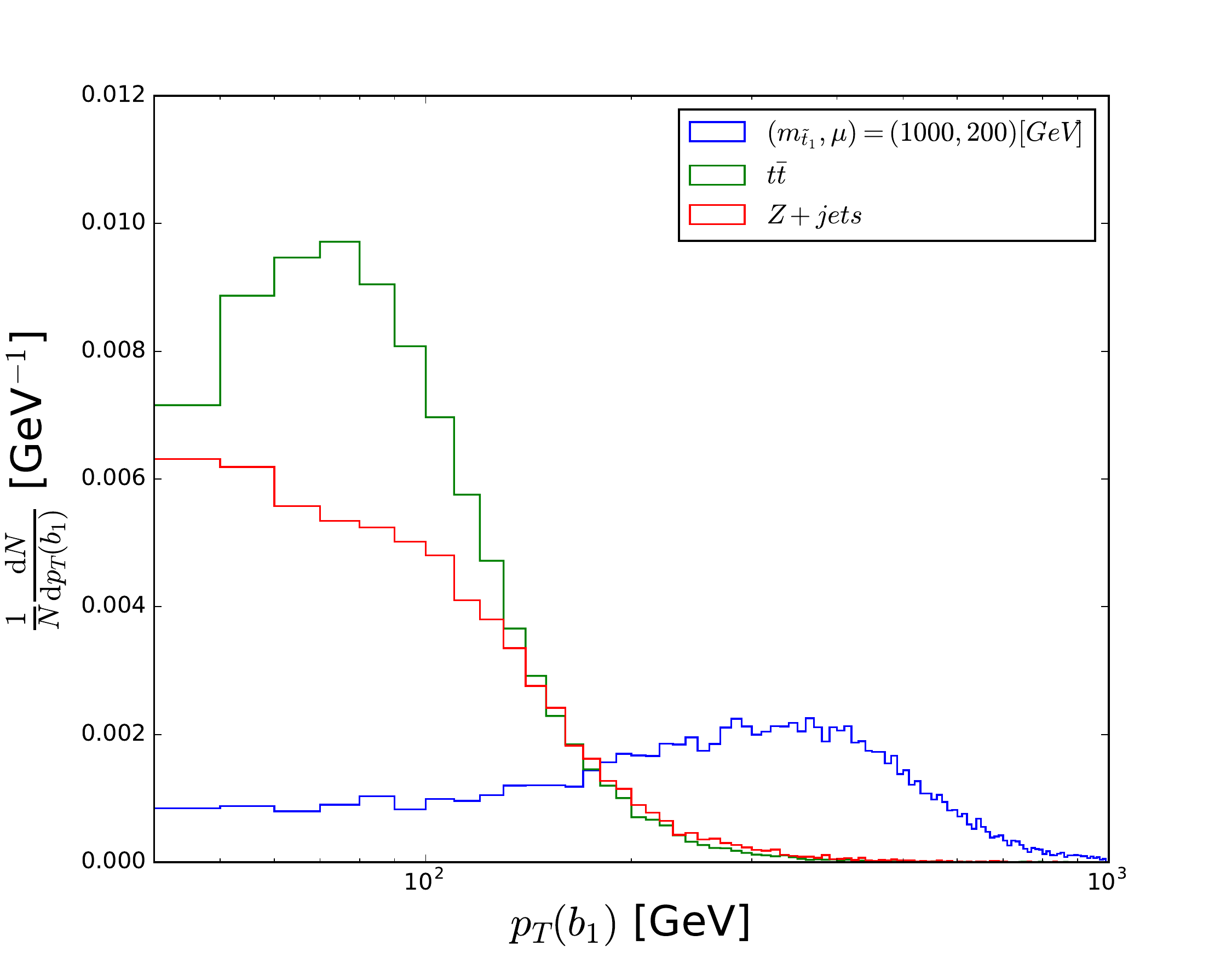}
\includegraphics[width=8cm,height=9cm]{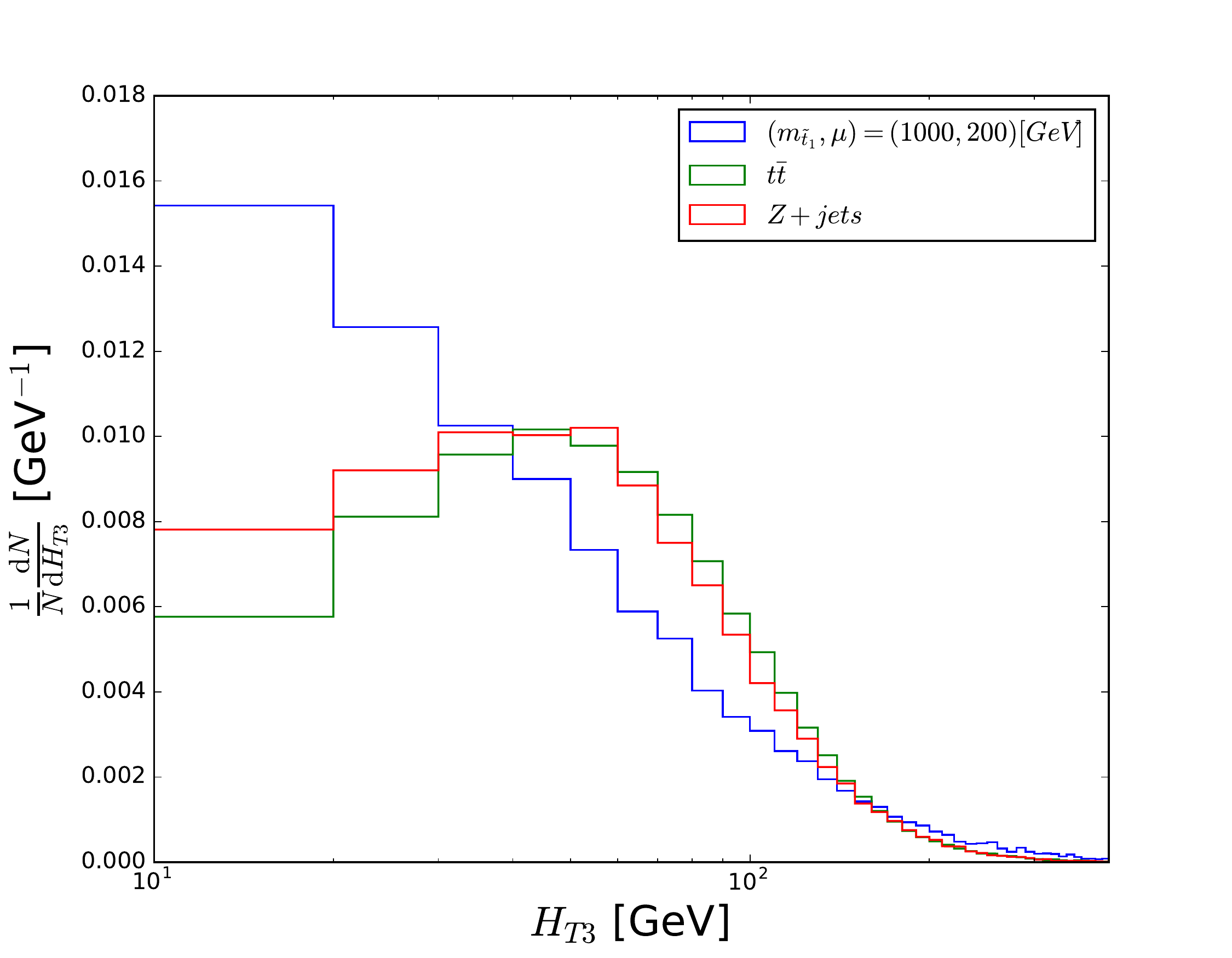}
\caption{The normalized distributions of $\slashed E_T$, $N(b)$, $p_{T}(b_{1})$ and $H_{T3}$ for the signal and the background events at 14 TeV LHC. The benchmark point is $m_{\tilde{t}_1}=1000$ GeV and $\mu=200$ GeV.}
\label{fig:distribution_14TeV}
\end{figure}
%%%--end--%%%

%%%--distribution_27TeV--%%%
\begin{figure}[htbp]
%\small
\centering
\includegraphics[width=8cm,height=9cm]{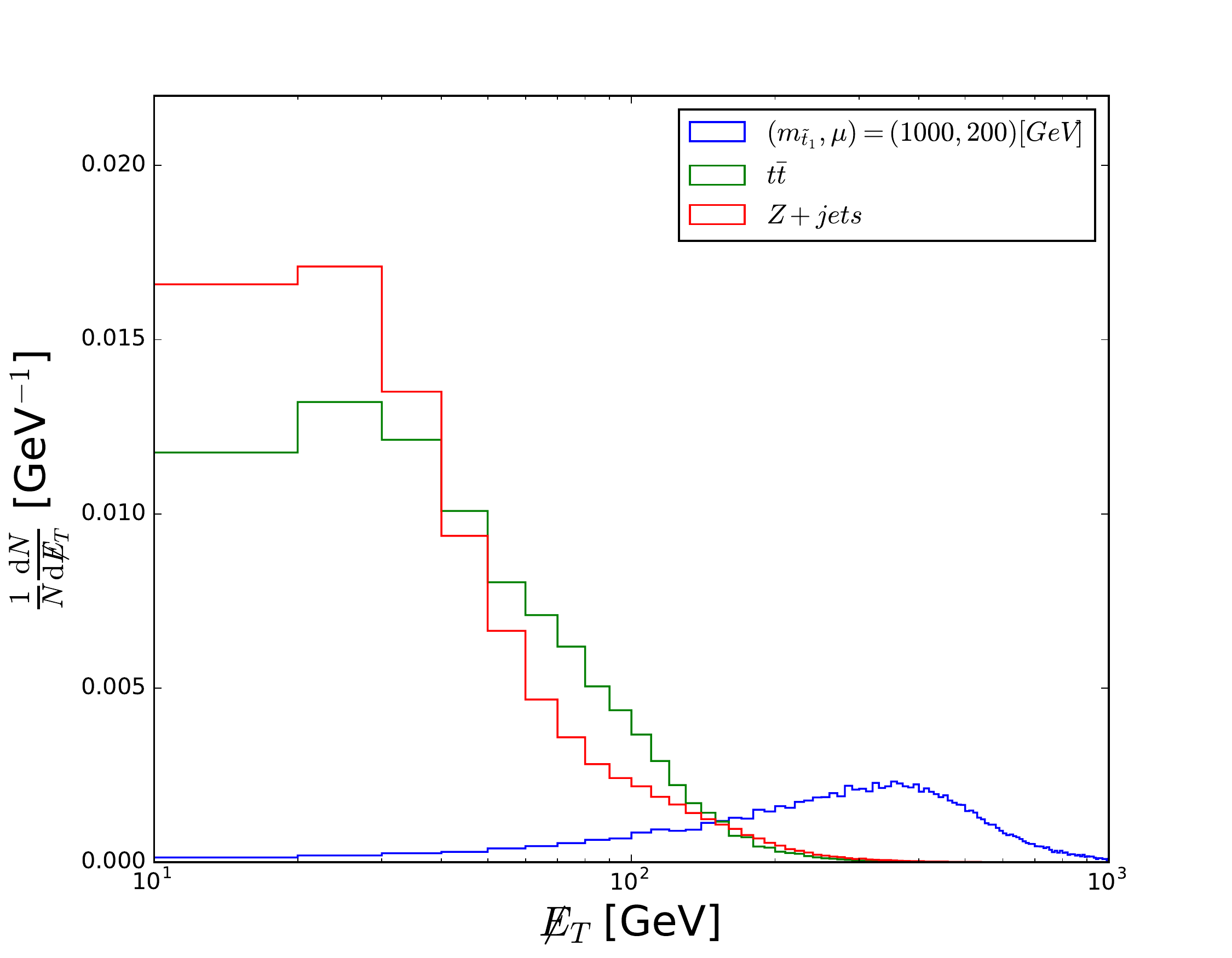}
\includegraphics[width=8cm,height=9cm]{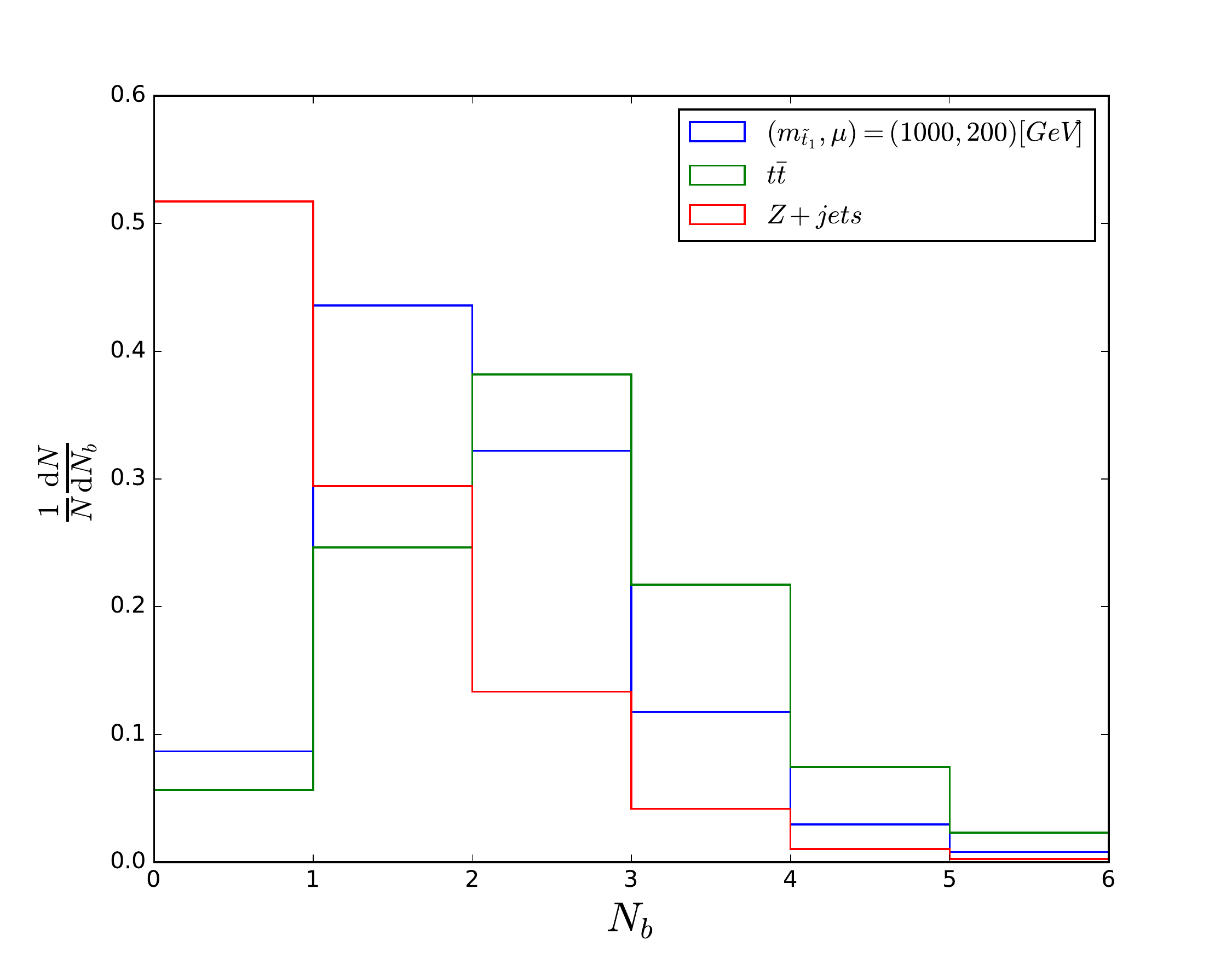}
\includegraphics[width=8cm,height=9cm]{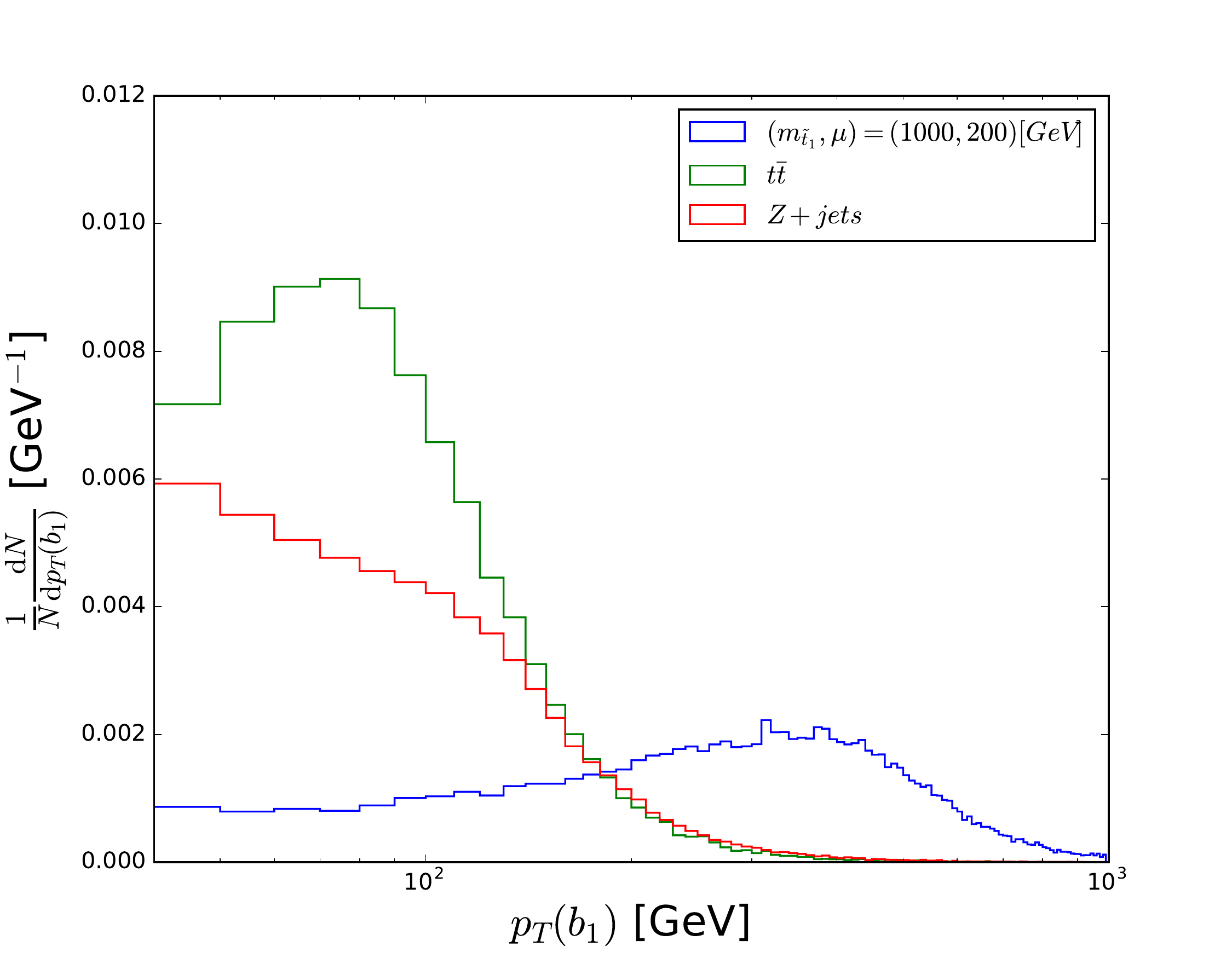}
\includegraphics[width=8cm,height=9cm]{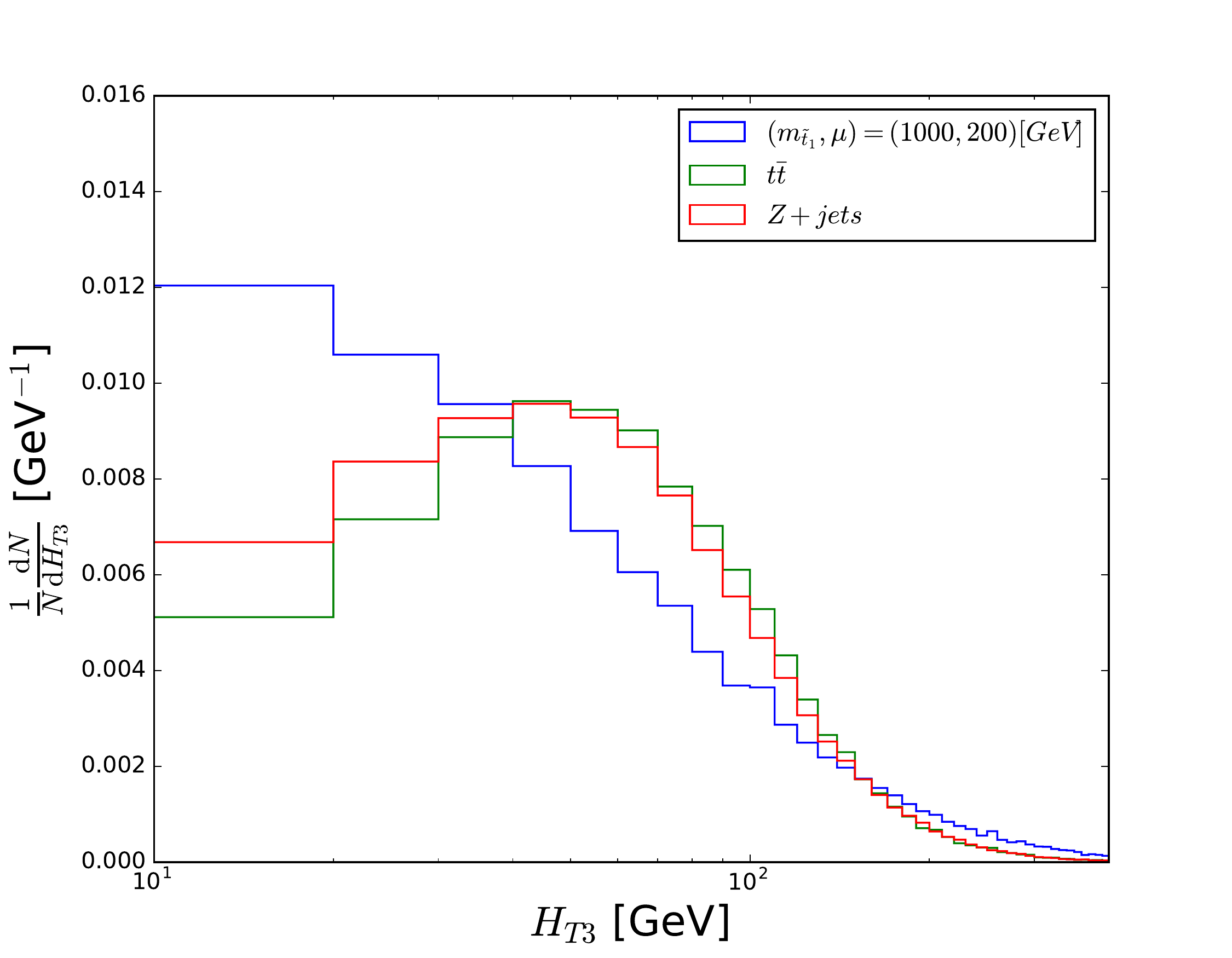}
\caption{Same as Fig.~\ref{fig:distribution_14TeV}, but for the HE-LHC.}
\label{fig:distribution_27TeV}
\end{figure}
%%%--end--%%%
In Fig.~\ref{fig:distribution_14TeV}, we show the normalized distributions of $\slashed E_T$ (the transverse missing energy), $N(b)$ (the number of $b$ jets), $p_T(b_1)$ (the transverse momentum of the leading $b$-jet) and $H_{T3}$ (the scalar sum of the transverse momentum of the third to fifth jet)~\cite{HT3} for the signal and the background events at the 14 TeV LHC. The signal events has a larger $\slashed E_T$ due to the massive $\tilde{\chi}^\pm_1$. The signal events have a larger $p_{T}(b_{1})$ because the $b$-jet from the stop decay is boosted. The majority of $Z+jets$ and $t\bar t$ background events are distributed in the region of $\slashed E_T \lesssim 350$ GeV and $p_{T}(b_{1}) \lesssim 400$ GeV. Besides, the $Z$+jets events have the least $b$-jets in the final states.
%{\color{red}It should be mentioned that, our background $Z+\text{jets}$ actually includes the Higgs-mediated process $pp\rightarrow Zh(\rightarrow b\bar{b})$. When one of the bottom quarks from Higgs decay is mis-identified, this process ends up with $b+\slashed{E}_{T}$ and hence constitutes our background. But in the events selection below, one of our cuts for $p_{T}(b)$ is $p_{T}(b_{1})>500\,(550)$ GeV for HL-LHC\,(HE-LHC), which will automatically filter out the above Higgs-mediated process in the background. Thus, the distributions for background $Z+\text{jets}$ that we present here are from events $pp\rightarrow Z+\text{jets}$ without the Higgs-mediated process.}
The $t\bar t$ background events can be separated from the signal events in the $H_{T3}$ distribution since there are fewer hard jets in the signal events. Similarly, these distributions are also shown for the HE-LHC in Fig.~\ref{fig:distribution_27TeV}, where the signal events have larger $\slashed E_T$ and $p_T(b_1)$ than the background events.

In our analysis, we perform the event selections as follows:
\begin{itemize}
  \item The events with any isolated leptons are rejected.
  \item We require at least two jets, and at least one $b$-jet with the leading $b$-jet $p_{T}(b_{1}) > 500$ GeV at the HL-LHC and $p_{T}(b_{1}) > 550$ GeV at the HE-LHC.
  \item We require $\slashed E_T > 450$ GeV at the HL-LHC and $\slashed E_T > 500$ GeV at the HE-LHC.
  \item $H_{T3} < 100$ GeV at the HL-LHC and $H_{T3} < 150$ GeV at the HE-LHC are required to further suppress the top pair background events.
  \item A minimum azimuthal angle between any of the jets and the missing transverse momentum $\Delta \phi(j,\slashed {\vec{p}}_T) > 0.6$ is required to reduce the multi-jet background.
\end{itemize}
With the above cuts applied, we can remove a large amount of background events and keep as many signal events as possible, for both the HE-LHC and HL-LHC.

\begin{figure}[bt]
%\small
\centering
\includegraphics[width=8cm, height=9cm]{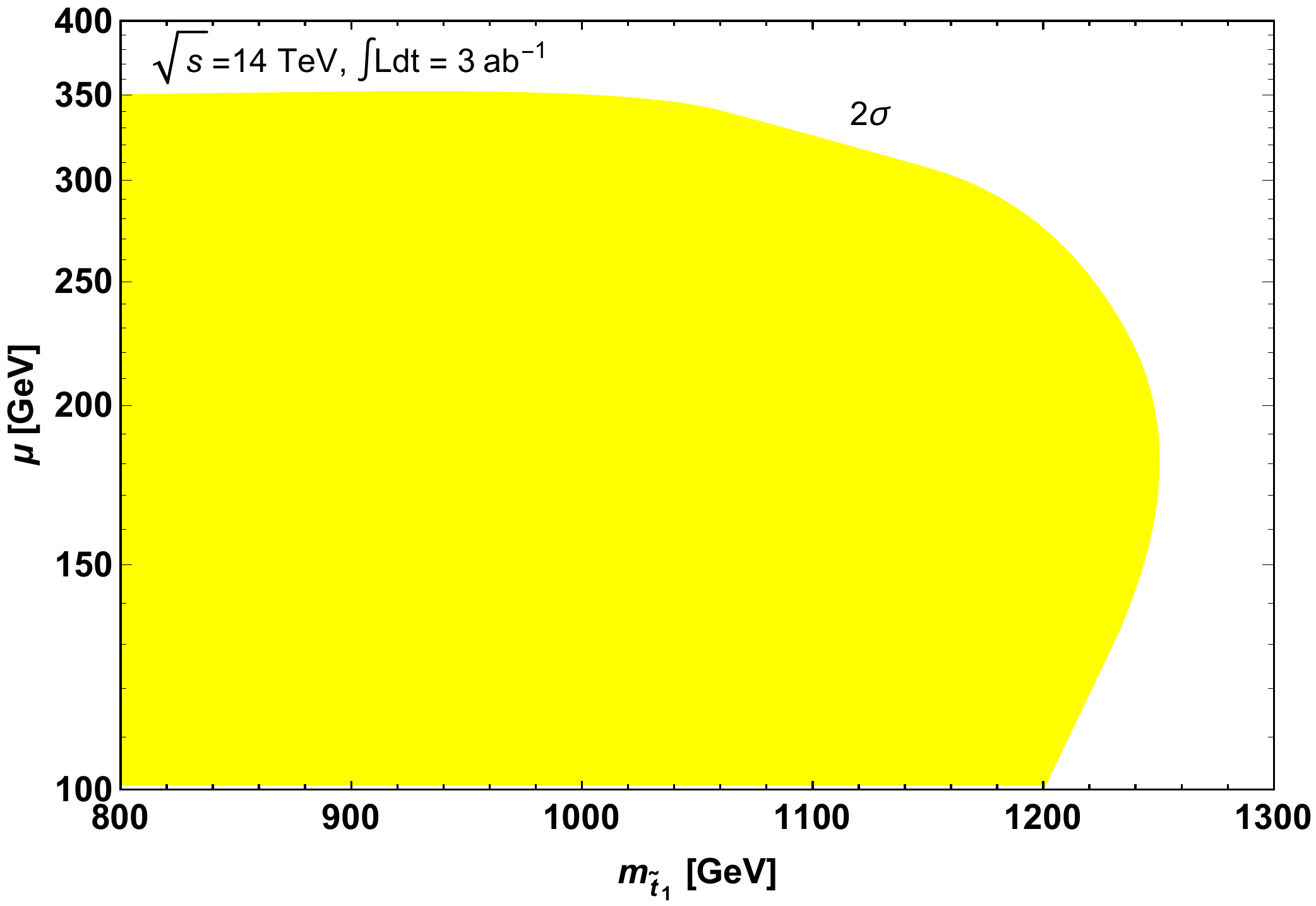}
\includegraphics[width=8cm, height=9cm]{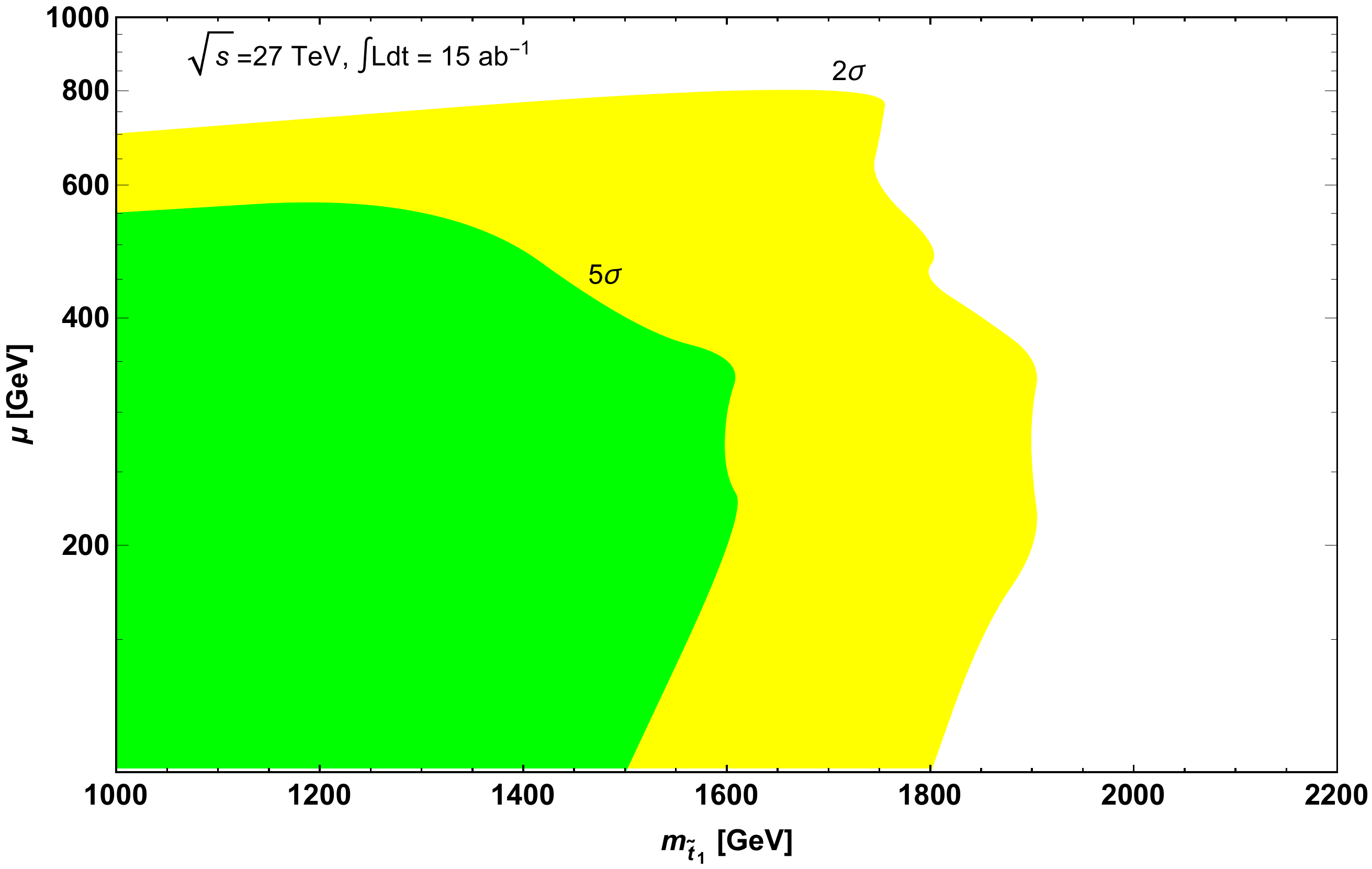}
\caption{The statistical significance $S/\sqrt{B}$ of the process $pp \to \tilde{t}_1 \tilde{\chi}^-_1 \to b+\slashed E_t$ on the plane of stop mass $m_{\tilde{t}_{1}}$ versus the higgsino mass parameter $\mu$ at the HL-LHC and HE-LHC.}
\label{fig:contour14}
\end{figure}

In Fig.~\ref{fig:contour14}, we present the contour plot of the statistical significance $S/\sqrt{B}$ of the process $pp \to \tilde{t}_1 \tilde{\chi}^-_1 \to b+\slashed E_t$ on the plane of stop mass $m_{\tilde{t}_{1}}$ versus the higgsino mass parameter $\mu$ at the HL-LHC and HE-LHC. We find that the stop mass $m_{\tilde{t}_{1}}$ and the higgsino mass $\mu$ can be excluded up to roughly 1.25 TeV and 350 GeV at $2\sigma$ level at the HL-LHC, respectively. Such a bound will be extended by about 650 GeV for the stop mass and 400 GeV for the higgsino mass at the HE-LHC. On the other hand, the HE-LHC will be able to cover the stop with the mass $m_{\tilde{t}_{1}} \lesssim 1.6$ TeV and the higgsinos with the mass $\mu \lesssim 550$ GeV at $5\sigma$ level. However, it should be mentioned that, due to the small values of $S/B$, our estimation of the statistical significance will become degraded when systematic uncertainties are taken into account. In order to reach the expected sensitivity, we need to control the systematic uncertainty at a few percent level. The accurate estimation of the systematic uncertainties relies on the real performance of the future detectors and is beyond the scope of this paper. Besides, we expect our analysis can be improved by advanced signal extraction strategies (such as machine learning technique~\cite{Albertsson:2018maf,Abdughani:2019wuv,Ren:2017ymm,Caron:2016hib}) and better understanding of the backgrounds uncertainties through the dedicated analysis of the experimental collaborations at future hadron colliders.

%%%--contour14 and 27--%%%
%%%--end--%%%

Again, we should point out that the single stop production may not be a discovery channel of stop. However, if the stop was discovered in the stop pair channel, then the single stop production will be helpful to identify the underlying theory and learn more about the properties of the sparticles. For example, in the AMSB supersymmetric model, the lightest chargino $\tilde{\chi}^-_1$ is usually wino-like, while in the Focus Point supersymmetric model it is preferred to be higgsino-like. The cross section of the single stop production process $pp \to \tilde{t}_1 \tilde{\chi}^-_1$ in the latter case is larger than that in the former case for the same stop and chargino masses~\cite{wu-2}. Besides, in the Natural SUSY, the stop can be right-handed or left-handed and $\tilde{\chi}^-_1$ is higgsino-like, which predicts that the production rate of $\tilde{t}_R\tilde{\chi}^-_1$ is larger than that of $\tilde{t}_L\tilde{\chi}^-_1$~\cite{wu-1}. Therefore, we conclude that it is meaningful to study the electroweak production of the stop, even though the stop is still not observed at this stage. After the discovery of stop, the future precision measurement of the cross section of the process $pp \to \tilde{t}_1 \tilde{\chi}^-_1$ could be used to distinguish different supersymmetric models, or even identify the nature of stop and electroweakinos.

\section{CONCLUSION}
In this work, we studied the single stop production $pp \to \tilde{t}_1 \tilde{\chi}^{-}_{1}$ in a simplified MSSM framework where the higgsinos and right-handed stop are the only sparticles. Different from the conventional $t\bar{t}+\slashed E_T$ signature of the stop pair production, the single stop production predicts some distinctive signatures at the HL-LHC and HE-LHC, which will be useful to confirm the existence of the stop. We analyzed the sensitivity of the mono-$b$ events from the single stop production process $pp \to \tilde{t}_1 \tilde{\chi}^{-}_{1} \to b \tilde{\chi}^+_{1} \tilde{\chi}^-_1 \to b+ \slashed E_T$. If the systematic uncertainty can be reduced to a few percent level, we found that the mass reach of the stop may be up to about 1.6 TeV at $5\sigma$ statistical significance at the HE-LHC with the integrated luminosity ${\cal L}=15$ ab$^{-1}$. In addition, if there was no significant excess in such a channel, the stop mass may be excluded up to about 1.25 TeV at the HL-LHC and 1.9 TeV at the HE-LHC.

\section*{Acknowledgement}
T. Tang would like to thank Murat Abdughani, Jie Ren and Jun Zhao for helpful discussions. This work was supported by the National Natural Science Foundation of China (NNSFC) under grants No.\,11705093 and No.\,11847208, and by the Jiangsu Planned Projects for Postdoctoral Research Funds, Grant No.\,2019K197.

\end{document}